\documentclass[english]{achemso}
\usepackage[T1]{fontenc}
\usepackage{amsmath}
\usepackage{graphicx}
\setkeys{acs}{usetitle = true,doi = true}

\makeatletter

\title{Ion Conductivity and Correlations in Model Salt-Doped Polymers: Effects
of Interaction Strength and Concentration}

\author{Kuan-Hsuan Shen}

\affiliation{William G. Lowrie Department of Chemical and Biomolecular Engineering,
The Ohio State University, Columbus, Ohio 43210, United States}

\author{Lisa M. Hall}

\affiliation{William G. Lowrie Department of Chemical and Biomolecular Engineering,
The Ohio State University, Columbus, Ohio 43210, United States}

\email{hall.1004@osu.edu}

\usepackage{color, soul}

\makeatother

\usepackage{babel}
\begin{document}
\begin{abstract}
Correlated anion and cation motion can significantly reduce the overall ion conductivity in electrolytes versus the ideal conductivity calculated based on the diffusion constants alone. Using coarse-grained molecular dynamics simulations, we calculate conductivity and the degree of uncorrelated ion motion in salt-doped homopolymers and block copolymers as a function of concentration and interaction strengths. Calculating conductivity from ion mobility under an applied electric field increases accuracy versus the typical use of fluctuation dissipation relationships in equilibrium simulations. In typical electrolytes, correlation in cation-anion motion is often expected to be reduced at low ion concentrations. However, for these polymer electrolytes with strong ion-polymer and ion-ion interactions, we find correlations are increased at lower concentrations when other variables are held constant. We show this phenomenon is related to the slower ion cluster relaxation rate at low concentrations rather than the static spatial state of ion aggregation or the fraction of free ions. 
\end{abstract}

\section{Introduction}

Ion-containing polymers are attractive materials for use as battery electrolytes
due to their mechanical robustness, electrochemical stability, and
low flammability. Their ion conductivity and other material properties
depend on a variety of experimentally tunable parameters such as polymer dielectric
strength, architecture, and ion types.\cite{ketkar2019charging,miller2017designing}
The diffusion of ions as a function of polymer segmental dynamics,
solvation site connectivity, and other factors has been extensively
studied.\cite{ganesan2012mechanisms,webb2015systematic,cheng2018design,shen2018diffusion,seo2019role}
However, since the transport
of neutral ion clusters does not contribute to energy generation,
correlation in cation and anion motion must also be understood to
fully capture ion conductivity. It has been shown that conductivity in polymer electrolytes
can vary significantly from the Nernst-Einstein
limit, which is based solely on ion density and ions' self-diffusion
constants.\cite{gainaru2016mechanism,stacy2018fundamental,galluzzo2020measurement}

Correlation in ion motion has been studied using various types of
experimental methods,
and different definitions and terminologies are used to describe
features of polymer electrolytes related to their correlated ion
motion and conductivity. In particular, the degree of ion
association/dissociation\cite{colby1997polyelectrolyte,timachova2015effect,buss2017nonaqueous,krachkovskiy2017determination},
``effective charge'',\cite{gouverneur2015direct,rosenwinkel2019lithium}
and Haven ratio (ratio between ions' diffusion constant and mobility,
equivalent to fractional deviation of conductivity from the Nernst-Einstein
equation)\cite{diederichsen2018investigation} are key properties that
can shed
light on ion correlation, yet do not directly correspond to each
other. The terms ion association or ion dissociation refer to the
spatial state of ionic aggregation (a structural property),
whereas the latter two describe the extent to which the motion of
an ion affects another. In fact, results from this work and prior
simulations have shown that analyzing structural ionic
aggregation, without considering the continual redistribution of ions,
does not allow one to completely
explain ion conductivity.\cite{fong2019ion,lonergan1995polymer,payne1995simulations,harris2010relations}

Many efforts are underway to better understand and quantify ion motion
and relate it to factors relevant to optimizing electrolyte performance.
Experimentally, a cluster of studies focused on transference
number ($t_{+}$, fraction of conductivity contributed by the cation)
and showed that ideal $t_{+}$ calculated from self-diffusion constants,
based on pulsed-field-gradient nuclear magnetic resonance (NMR) measurements, significantly
differs from $t_{+}$ calculated from approaches with correlated ion
motion taken into account.\cite{pesko2017negative,gouverneur2018negative,villaluenga2018negative}
The difference between various definitions of $t_{+}$ is also evident in
computational work,\cite{molinari2019general,fong2019ion,molinari2019transport}
highlighting the importance of understanding correlation in ion motion.

Importantly, recent computational studies show that the fractional deviation
of conductivity from the Nernst-Einstein equation (denoted as $\Lambda/\Lambda_{NE}$
in the present work and often called the degree of uncorrelated
ion motion in simulation work) can be enhanced by increasing polymer dielectric
constant (polarity), which can improve the overall conductivity.\cite{wheatle2017influence,wheatle2018effect,wheatle2019influence}
Some work also investigated the separate contributions from the
various types of cross correlation terms to the total ionic conductivity
(overall correlations can be divided into cation-cation, anion-anion,
and cation--anion correlations).\cite{kashyap2011charge,fong2019ion,zhang2019ion}

Despite the abundance of computational work on conductivity, accurate
and efficient measurement of conductivity from molecular dynamics
simulations remains challenging. Conductivity can be
directly calculated, including effects of correlated ion motion, from
equilibrium simulations with the use of fluctuation dissipation relationships.
However, this approach typically introduces a large statistical uncertainty
from measuring the collective displacement of ions. Unphysical overall
$\Lambda/\Lambda_{NE}$ values larger than $1.0$ were sometimes
encountered, apparently
due to the poor statistics.\cite{borodin2006mechanism,wheatle2018effect,wheatle2019influence,wheatle2020effect}
Several methods have been used to reduce statistical noise in the conductivity
calculation such as 1) using only the short-time scale data to allow
for significant averaging and because the
collective ion migration terms tend to fluctuate wildly at long time
scales,\cite{borodin2006mechanism,seo2019role,wheatle2017influence,wheatle2018effect,wheatle2019influence,wheatle2020effect}
2) considering ion clusters as noninteracting charge carriers and
applying the Nernst-Einstein equation with the diffusion constants and
net charges of the different types of ion clusters,\cite{france2019correlations} and 3)
measuring ion mobility from nonequilibrium simulations with an external
electric field.\cite{ting2015structure,weyman2018microphase,alshammasi2018correlation,wheeler2004molecular2,zhang2019ion}
Prior studies have compared conductivities estimated by different
methods for several specific systems.\cite{wheeler2004molecular1,wheeler2004molecular2,liu2011molecular}
Here, we present a side-by-side analysis of ion correlations in a
variety of systems
calcuated by nonequilibrium versus equilibrium methods, including
those in a control system with $\Lambda/\Lambda_{NE}$=1, to
clearly assess the accuracy and efficiency of these methods.

In our previous work, we applied our newly-developed model (with a
$1/r^{4}$ potential form to represent ion solvation) to study the
effect of molecular weight on ion diffusion constant in salt-doped homopolymer
and block copolymer electrolytes.\cite{seo2019role} In this study,
we apply the same model to probe ion conductivity
and ion correlation in both salt-doped homopolymer and block copolymer
systems. We note that we focus on $\Lambda/\Lambda_{NE}$ to study
correlated ion motion in this work, and the transference number will
be probed in further work. 

First, we measured $\Lambda/\Lambda_{NE}$ in toy systems of salt-doped
polymers in which no Coulomb potential between ions was applied, using
both equilibrium and nonequilibrium methods. In particular, for the nonequilibrium method, we applied
an electric field in one direction (aligned along lamellae in the case
of block copolymers), which allows us to directly calculate
both ion mobility and diffusion constant from ions' displacement parallel
and perpendicular to the electric field, respectively. We ensured
the field is low enough that the systems are in the linear response
regime while still allowing for mobility high enough to measure accurately
in the timescale of the simulation. Then, we probed how $\Lambda/\Lambda_{NE}$
relates to Coulomb strength and ion concentration. Fraction of free
ions, polymer, and cluster relaxation rates were also calculated to
help explain the trends in ion motion correlation. Finally, we studied
the effects of ion-monomer and ion-ion solvation strengths on ion
correlation and ion conductivity at various amounts of salt loading. By elucidating
how these key factors affect ion conductivity, we hope to suggest
design rules to optimize conduction in future materials.

\section{Methods}

\subsection{Simulation Model}

The coarse-grained model we use is based on that of Refs \citenum{seo2019role, brown2018ion},
which includes standard Kremer-Grest bead-spring chains with equal amounts
of anions and cations added.\cite{grest1996efficient} Both salt-doped
homopolymer (HP) and block copolymer (BCP) systems are studied, where
each chain has a total of 20 and 40 monomer beads, respectively. Homopolymer
chains consist of only type A monomer beads, giving a fraction of
A monomers ($f_{A}$) of 1. Block copolymer chains contain equal amounts
of two monomer types, A and B, with $f_{A}=0.5$. Bonded monomers are
subject to the finitely extensible nonlinear
elastic (FENE) potential:
\begin{equation}
U_{FENE}(r)=-0.5kR_{0}\ln\left(1-\frac{r^{2}}{R_{0}^{2}}\right)
\end{equation}
in which $r$ is the distance between the two monomers and the spring constant
$k=30\epsilon/\sigma^{2}$ and the maximum distance between two bonded
beads $R_{0}=1.5$ are standard values chosen to avoid chains crossing each other.
All beads interact via the Leonard-Jones (LJ) potential:

\begin{equation}
U_{\textrm{LJ,ij}}(r)=\begin{cases}
4\epsilon\Bigg[\bigg(\dfrac{\sigma_{ij}}{r}\bigg)^{12}-\bigg(\dfrac{\sigma_{ij}}{r}\bigg)^{6}-\bigg(\dfrac{\sigma_{ij}}{r_{c}}\bigg)^{12}+\bigg(\dfrac{\sigma_{ij}}{r_{c}}\bigg)^{6}\Bigg] & r\leq r_{c}\\
0 & r>r_{c}
\end{cases}
\end{equation}
where the cutoff distance $r_{c}=2^{1/6}\sigma$. All beads have a
diameter of $\sigma_{ij}=1.0\sigma$ and unit mass. The LJ interaction
strength $\epsilon_{ij}=1.0\epsilon$ for all interactions other than
A-B interactions, and $\epsilon_{AB}=2.0\epsilon$ so that the interaction
is unfavorable enough for the block copolymer system to microphase
separate.\cite{seo2019role} A wide range of ion concentration were
tested, reported as the number ratio of cations to A monomers $[+]/[A]=0.006$,
$0.013$, $0.026$, $0.052$, $0.104$, $0.156$, and $0.208$. Ion-ion
interactions include the Coulomb potential:

\begin{equation}
U_{Coulomb,ij}(r)=\frac{q_{i}q_{j}}{4\pi\epsilon_{0}\epsilon_{r}r}
\end{equation}
where $q_{i}$ and $q_{j}$ are individual charges of the interacting
ion pairs ($+1e$ or $-1e$), $\epsilon_{0}$ is the vacuum permittivity,
and $\epsilon_{r}$ is the dielectric constant of the medium. We report
Coulomb strength in terms of the Bjerrum length $l_{B}=e^{2}/(4\pi\epsilon_{0}\epsilon_{r}k_{B}T)$,
the distance between two electron charges at which their electrostatic
potential is comparable in magnitude to the thermal energy scale
$k_{B}T$. We report results in standard reduced LJ units (lengths are
in units of the monomer diameter $\sigma$ and energy is in units of
$k_{B}T$=$\epsilon$, the LJ interaction strength between same type monomers).
To map to real experimental systems as reference, we can follow the same approach as in Refs \citenum{seo2019role, brown2018ion}:
we consider polyethylene oxide (PEO) at $T=400K$ as the conducting phase, which has a
dielectric constant $\epsilon_{r}\approx7.5$.\cite{nakamura2011thermodynamics,porter1971dielectric,gray1989study} With $1.0\sigma$ (the
contact distance between a cation and an anion in our simulations)
being mapped to $0.7nm$ (the distance from Li$^{+}$ to the center
of a bis(trifluoromethylsulfonyl)imide anion (TFSI$^{-}$) in PEO),\cite{mao2000structure}
the Bjerrum length $l_{B}\approx8\sigma$. Below, $l_{B}=8$ is used for all the simulations
unless otherwise noted. To explore the Coulomb strength effects
on ion dynamics, we also simulate systems with $l_{B}=0\sigma$, $4\sigma$,
and $12\sigma$. In the case of $l_{B}=0\sigma$, the Coulomb
interaction between ions is not applied, however, they are still
identified as cations or anions by their ``charge'' and we analyze
them as such in calculations. The dielectric constant is the same throughout the
simulation box even for microphase separated block copolymer systems;
effectively, we assume ions interact as though they are only in the
conducting phase, given that in the experimental systems of interest they are strongly segregated to the conducting
phase. As motivated and described in our prior work, additional ion-monomer and ion-ion
solvation interactions that drive this segregation to the conducting phase are also applied:

\begin{equation}
U_{Solvation,ij}(r)=\begin{cases}
-S_{ij}\Bigg[\bigg(\dfrac{\sigma}{r}\bigg)^{4}-\bigg(\dfrac{\sigma}{r_{c}}\bigg)^{4}\Bigg] & r\leq r_{c}\\
0 & r>r_{c}
\end{cases}
\end{equation}
where the cutoff distance $r_{c}=5\sigma$. The solvation strength
$S_{ij}$ is always the same between certain types of beads: $S_{AA}=S_{AB}=S_{BB}=0\epsilon$,
$S_{A+}=S_{A-}=S_{A\pm}$, $S_{B+}=S_{B-}=S_{B\pm}$, and $S_{++}=S_{+-}=S_{--}=S_{\pm\pm}$,
with the $+$ subscript referring to cations and the $-$ to anions.
The solvation strength difference between A and B phases is denoted
as $\Delta S=S_{A\pm}-S_{B\pm}$. This difference is expected to be
the relevant factor in setting the structure of the block copolymer
system, rather than either ion-monomer solvation parameter alone. Based on the
dielectric constant of the host polymer and ion size (which is $1\sigma$ throughout the work), the Born solvation
energy $\Delta V_{Born}$ and the corresponding solvation parameter can be calculated
as in Refs \citenum{seo2019role, brown2018ion}. Briefly, using $\Delta
V_{Born}=[e^{2}/(4\pi\epsilon_{0}\sigma)](1/\epsilon_{r}-1)$, the Born
solvation energy is $-ˆ'52k_{B}T$ for PEO with a
dielectric constant of 7.5,\cite{nakamura2011thermodynamics,porter1971dielectric,gray1989study} $-46k_{B}T$ for polycaprolactone (PCL) with a dielectric
constant of 4.4,\cite{hegde2016dielectric} and $-36k_{B}T$ for polystyrene (PS) with a
dielectric constant of 2.5.\cite{roberts1946new,yano1971dynamic} The values of $S_{A\pm}$ (with PEO
as type A) and $S_{B\pm}$ (with PS or PCL as type B) in our simulations can then be obtained by approximating
the insertion energy of an ion into a pure A or B homopolymer system
relative to the insertion energy of the same ``ion'' but with no solvation
energy term, $S=0$. Each insertion energy is the integral of $\rho\int g_{(A/B)\pm}(r)u_{(A/B)\pm}(r)dr$ over
all space, where $g_{(A/B)\pm}(r)$ is the pair correlation function and $u_{(A/B)\pm}(r)$
is the total pairwise interaction potential. With the density and $g_{(A/B)\pm}(r)$ obtained from homopolymer systems ($N=20$) with low ion concentration, we
find $S_{A\pm}\approx4.4$ for PEO, $S_{B\pm}\approx4.0$ for PCL,
and $S_{B\pm}\approx3.2$ for PS. Thus, $\Delta S\approx0.4$ for
PCL-$b$-PEO and $\Delta S\approx1.2$ for PS-$b$-PEO. Since it is
not the goal of this work to match exactly the chemical details of
certain systems, we do not directly apply the above solvation parameters
for computational efficiency, allowing us to simulate a wide
range of systems to better understand the parameter space (system dynamics can be significantly
slowed down due to strong ion solvation, as shown in the Supporting
Information). Instead, we scaled down the parameters but kept large
enough $S_{A\pm}$ for correct ion diffusion trend as well as large
enough $\Delta S$ so that ions are mostly segregated to the A phase
(see Supporting Information for more detail). We also kept $S_{A\pm}=S_{\pm\pm}$
to avoid the diluent effect (the addition of ions
inherently dilutes the system) that can potentially lead to incorrect ion diffusion trend in block copolymer system at high
concentrations. Such effect can come
into play at high ion concentration when ion-ion interactions
become significant, if these are less favorable than ion-A interactions.
Here, we use $S_{A\pm}=S_{\pm\pm}=2.7$, $S_{B\pm}=2.3$ ($\Delta S=0.4$)
for all systems unless stated otherwise. To probe how the solvation
strengths impact ion transport, we also tested systems with $S_{A\pm}=2.0-3.5$,
$S_{\pm\pm}=0-3.0$, and $\Delta S=0.8-1.2$. 

The LAMMPS simulation software is used for all the simulations in
this study.\cite{plimpton1995fast} Snapshots are visualized with the VMD
software.\cite{humphrey1996vmd} Following the same procedure from
our previous studies,\cite{seo2017diffusion,shen2018diffusion,seo2019role}
homopolymers were initialized as random walks, while block copolymer
systems were initialized as random walks within the constraint of a
lamellar morphology. Ions were randomly added in
the whole simulation box for homopolymer systems or in the A phase
for the block copolymers. Initial overlap between monomers was eliminated
using a short soft push-off phase preceding equilibration. 
The simulations were then equilibrated in a NPT ensemble at temperature
$T=1.0\epsilon/k_{B}$ and pressure $P=5.0\epsilon\sigma^{-3}$ for
$5\times10^{4}\tau$ using Nos\'e-Hoover thermostat and barostat, both
with a damping parameter of $5.0\tau$. For homopolymer systems, the barostat coupled the three dimensions of the box to keep a cubic box. For block copolymer systems, the barostat coupled x and y box dimensions
to each other to keep a square cross section and allowed the z box
length (perpendicular to the lamellae) to fluctuate independently so that each system could reach its
equilibrium lamellar domain spacing. We calculated polymers' mean squared displacement
(MSD) and ensured they had moved more than a few times their
radius of gyration on average; this is one measure that the polymers likely had explored various
conformations and approximately reached an equilibrium conformational
state. Subsequent to the NPT equilibration,
we fixed the box size using the average spacing and density obtained
from the last $5\times10^{3}\tau$ of the NPT run. The system was
then simulated in the NVT ensemble (under the same conditions except
with no barostat) for $5\times10^{4}\tau$. During this time the
systems reach the Fickian (diffusive) regime. The system was then simulated
for $6\times10^{5}\tau$ to collect data for dynamic analysis. A time
step of $\delta t=5\times10^{-3}\tau$ was used throughout this work. The simulation outputs were saved every $100\tau$ for structural analysis as well as for conductivity calculation, while for ion cluster and polymer relaxation rate analysis, data were saved at timesteps of every power of 2 starting from 2 and
3 such as $2^{1},3,2^{2},3\times2^{1},2^{3},3\times2^{2}$, etc.,
up to $\approx3940\tau$, at which point the logarithmic spacing repeated.

To compare conductivity and correlation results between equilibrium MD (EMD) and nonequilibrium MD (NEMD) simulations,
we ran 8 systems from different initial configurations at $l_{B}=0\sigma$
and $8\sigma$ and at various salt loadings. For NEMD simulations,
we apply an external, static electric field $E$ in the $x$-direction
in units of $\sigma/(k_{B}Te)$,
which gives an additional force $q_{i}E\hat{x}$ on ion $i$ (this force
is applied even when Coulomb interactions are turned off for the
$l_{B}=0\sigma$ case). The
electric field was turned on at the beginning of the NVT run, allowing
the systems to be under the field for $5\times10^{4}\tau$ before
data collection. The thermostat in the direction of the
field was turned off as in prior studies.\cite{ting2015structure,sampath2018impact}
We note there is no extra equilibration procedure for NEMD systems
comparing to those of EMD, so the total equilibration time is the
same for both methods for a fair comparison.

For both EMD and NEMD simulations, the ion MSD was block
averaged over $2-12$ nonoverlapping trajectories depending on the
system dynamics (depending on whether breaking the trajectory into
shorter blocks still allows it to conform to the metric we set
below to ensure it is in the Fickian regime). The diffusion constants were obtained using the slope
from a least squares linear fit of the last decade of logarithmically spaced 
MSD(t) data (only data at every power of 2 starting from $200$, $300$, $500$, and $700\tau$ in each block were used for the fitting). For example, for the slowest systems in which the total
time of $6\times10^{5}\tau$ is split into only two blocks and averaged to
create data up to a time of $3\times10^{5}\tau$, the data from $3\times10^{4}-3\times10^{5}\tau$ is fit. The log-log slopes of all systems\textquoteright{}
MSD(t) curves are in range of 0.98 to 1.02. In addition,
for NEMD simulations, the ion drift velocity was block averaged and
fitted the same way. The log-log slopes of all systems\textquoteright{}
ion drift velocity versus time curves are in range of 1.96 to 2.04.
The calculation of these ion transport properties is detailed in the
next section.

\subsection{Ion Conductivity and Degree of Uncorrelated Ion Motion}

\subsubsection{Equilibrium Method}

To probe the ion correlations in EMD, both the Nernst-Einstein molar conductivity
$\Lambda_{NE}$ and the true molar conductivity $\Lambda$ were calculated.
We also have reported the Nernst-Einstein total conductivity $\lambda_{NE}$
as well as the true total conductivity $\lambda$. We note that $\Lambda_{NE}$
and $\Lambda$ reported in this work are calculated on a per-ion basis; they are not in per-mol units but are proportional to the actual ``molar'' conductivity. The
Nernst-Einstein molar conductivity can be calculated by
\begin{equation}
\Lambda_{NE}=\frac{\lambda_{NE}}{N_{ion}}=\frac{1}{2dN_{ion}Vk_{B}T}\lim_{t\to\infty}\frac{d}{dt}\sum_{i=1}^{N_{ion}}q_{i}^{2}\bigl\langle\Delta r_{i}(t)^{2}\bigr\rangle=\frac{e^{2}(D_{+}+D_{-})}{Vk_{B}T}\label{eq:cond_NE}
\end{equation}
where $V$ is the system volume, $N_{ion}$ is the total number of
ions, $\Delta r_{i}(t)=r_{i}(t)-r_{i}(0)$ is the displacement of
ion $i$ between time $t$ and time $0$, $\bigl\langle\Delta r_{i}(t)^{2}\bigr\rangle$
is the MSD (after subtracting the total system center of mass displacement)
of ion $i$ at time $t$, $d$ is the number of spatial dimensions over which the MSD
is considered, and $D_{+}$ and $D_{-}$ are the diffusion constants
of cations and anions, respectively. In EMD, $d=3$ for HP and $d=2$
for BCP (the z-direction is not taken into account as it is perpendicular
to the interface that hinders ion diffusion). The second equality
is based on the Einstein relation:
\begin{equation}
D=\frac{1}{2dN_{ion}}\lim_{t\to\infty}\frac{d}{dt}\sum_{i=1}^{N_{ions}}\bigl\langle\Delta r_{i}(t)^{2}\bigr\rangle\label{eq:diffusion_Einstein}
\end{equation}
where $D$ is the diffusion constant of both cations and anions; in
this work, $D_{+}=D_{-}=D$ as cations and anions are identical except that
they have opposite charge.

\begin{figure}
\includegraphics[scale=0.5]{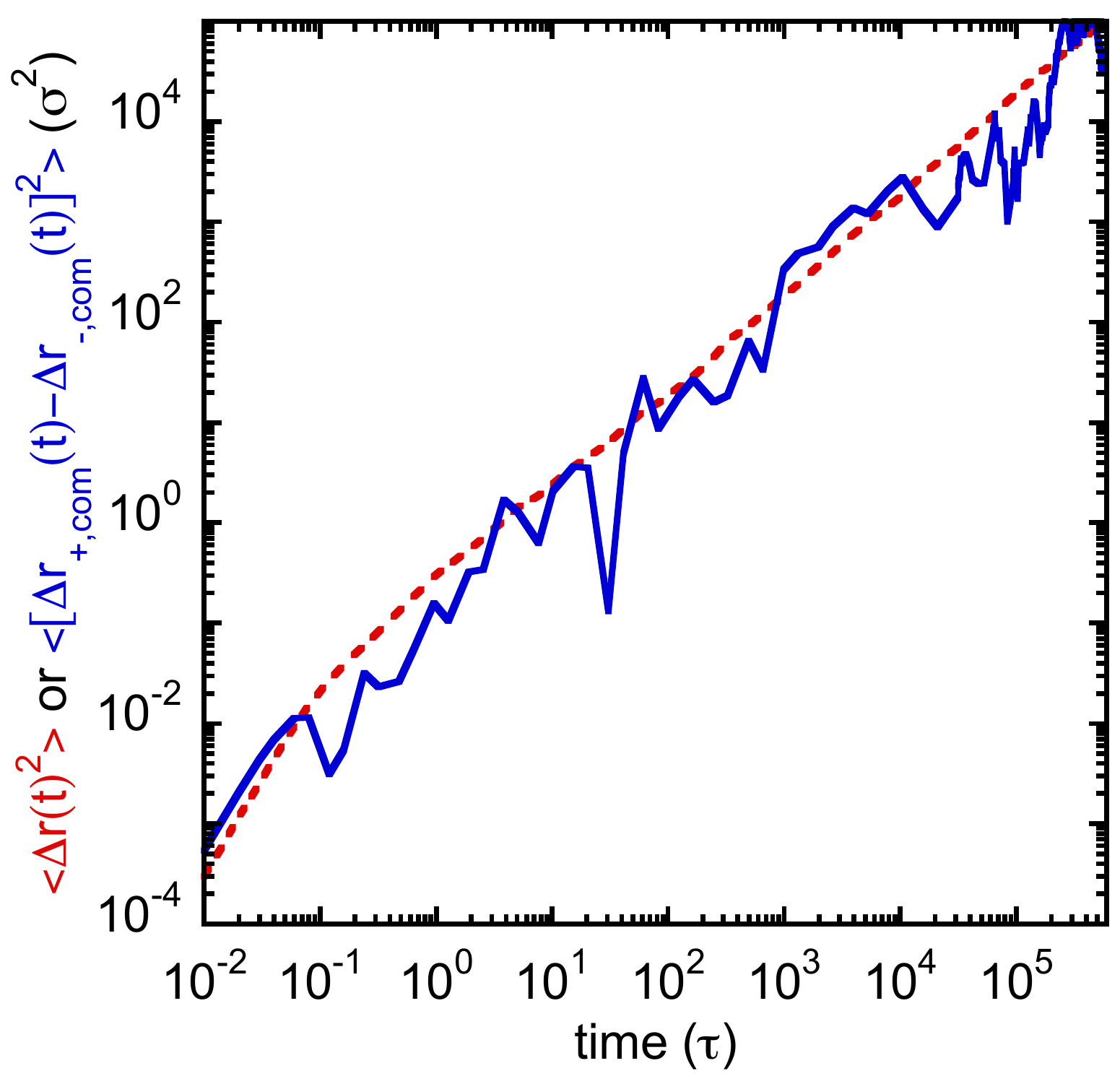}

\caption{The two components to calculate $\Lambda/\Lambda_{NE}$ in EMD: the
mean squared displacement of all ions (red dashed line) and the difference
in cations' and anion's center of mass displacement (blue solid line).
This example is from the homopolymer system at $l_{B}=0\sigma$ and $[+]/[A]=0.006$.\label{fig:EMD_msd}}
\end{figure}

The true molar conductivity can be calculated with collective ion
displacement taken into account using the fluctuation-dissipation
relation in EMD:\cite{hansen2013theory,harris2010relations}
\begin{equation}
\Lambda=\frac{\lambda}{N_{ion}}=\frac{1}{2dN_{ion}Vk_{B}T}\lim_{t\to\infty}\frac{d}{dt}\sum_{i=1}^{N_{ion}}\sum_{j=1}^{N_{ion}}q_{i}q_{j}\bigl\langle\Delta r_{i}(t)\Delta r_{j}(t)\bigr\rangle\label{eq:cond_EMD}
\end{equation}
Thus, the ratio $\Lambda/\Lambda_{NE}$ or the degree of uncorrelated
ion motion in EMD can be calculated as follows:
\begin{equation}
\left(\frac{\Lambda}{\Lambda_{NE}}\right)_{EMD}=\frac{\sum_{i=1}^{N_{ion}}\sum_{j=1}^{N_{ion}}q_{i}q_{j}\bigl\langle\Delta r_{i}(t)\Delta r_{j}(t)\bigr\rangle}{\sum_{i=1}^{N_{ion}}q_{i}^{2}\bigl\langle\Delta r_{i}(t)^{2}\bigr\rangle}=\frac{\Bigl\langle[\Delta r_{+,com}(t)-\Delta r_{-,com}(t)]^{2}\Bigr\rangle}{\bigl\langle\Delta r(t)^{2}\bigr\rangle}\label{eq:alpha_EMD}
\end{equation}
where $\bigl\langle\Delta r(t)^{2}\bigr\rangle$ is the MSD of all
ions; $\Delta r_{+,com}(t)$ and $\Delta r_{-,com}(t)$ are the center
of mass displacements of cations and anions, respectively. This ratio
represents the proportion of all the terms in the collective displacement
to only the self-correlation term. While the average of the self-correlation term is simply
the MSD, the average of the collective displacement can be simplified
as the difference in cations' versus anions' center of mass
displacement. It is thus intuitive that the collective term has larger
fluctuations than the self-correlation term, which involves a
squared quantitiy that is positive for each ion and averaged over all ions. 
As an representative example, the two terms needed to calculate $\Lambda/\Lambda_{NE}$
in EMD are individually plotted as a function of time in Figure \ref{fig:EMD_msd}.

\subsubsection{Nonequilibrium Method}

\begin{figure}[H]
\includegraphics[scale=0.5]{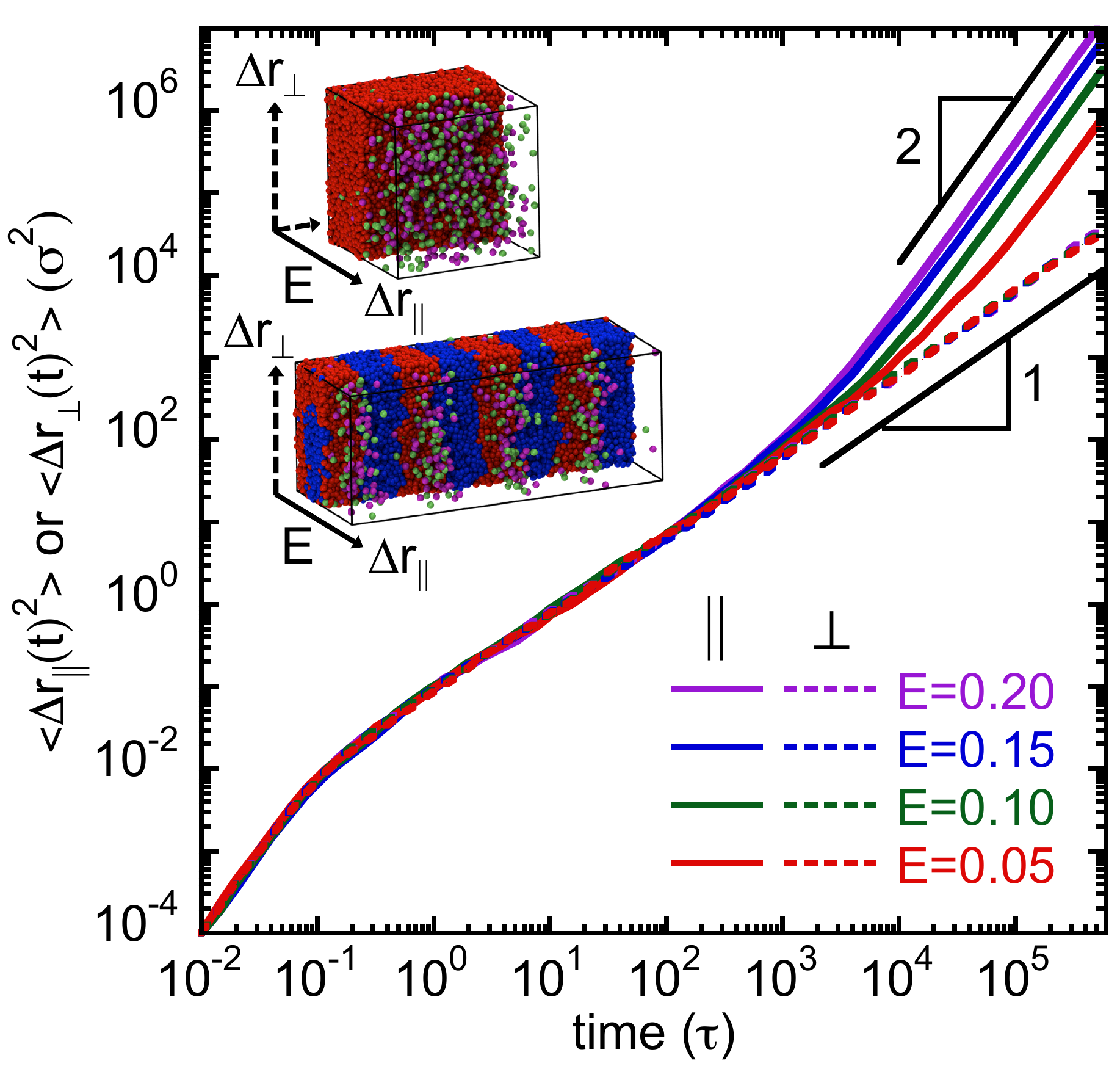}

\caption{The directional MSD in the parallel (solid lines) and perpendicular
(dashed lines) directions to the electric field with inset snapshots
showing the relative directions in homopolymer and block copolymer
systems. Two black lines indicate log-log slope of 1 and 2. These
examples are at $l_{B}=0\sigma$ and $[+]/[A]=0.006$ and the curves
shown are for the homopolymer system.
\label{fig:NEMD_msd}}
\end{figure}

We apply an external, static electric field $E$ in the x-direction.
At long times, ions have a steady state drift velocity along
the parallel direction of the field. The drift velocity can be calculated
from
\begin{equation}
\langle v_{x}\rangle^{2} t^{2}=\bigl\langle\Delta r_{x}(t)^{2}\bigr\rangle_{E}-\bigl\langle\Delta r_{x}(t)^{2}\bigr\rangle_{0}=\bigl\langle\Delta r_{\parallel}(t)^{2}\bigr\rangle_{E}-\bigl\langle\Delta r_{\perp}(t)^{2}\bigr\rangle_{E}
\end{equation}
where $\bigl\langle\bigr\rangle_{E}$ denotes an ensemble average
in NEMD where an electric field $E$ is applied in the x-direction,
$\bigl\langle\bigr\rangle_{0}$ denotes an ensemble average in EMD
at zero field; $\bigl\langle\Delta r_{\parallel}(t)^{2}\bigr\rangle_{E}$
and $\bigl\langle\Delta r_{\perp}(t)^{2}\bigr\rangle_{E}$ are the
directional MSDs of all ions in the parallel and perpendicular directions
to the field, respectively. Instead of using the first relation that
requires two simulations to be conducted, we calculated the drift
velocity with the second relation as the field has no effect on the
MSD in the perpendicular direction. Since the electric field is in
the x-direction, $\bigl\langle\Delta r_{\parallel}(t)^{2}\bigr\rangle_{E}$
is always equivalent to $\bigl\langle\Delta r_{x}(t)^{2}\bigr\rangle_{E}$
for all systems. While $\bigl\langle\Delta r_{\perp}(t)^{2}\bigr\rangle_{E}$
is the average y- and z-directional MSD for HP systems, i.e. $\bigl\langle\Delta r_{\perp}(t)^{2}\bigr\rangle_{E}=\left(\bigl\langle\Delta r_{y}(t)^{2}\bigr\rangle_{E}+\bigl\langle\Delta r_{z}(t)^{2}\bigr\rangle_{E}\right)/2$),
only the y-directional MSD (along the lamellae) is considered for BCP systems, i.e. $\bigl\langle\Delta r_{\perp}(t)^{2}\bigr\rangle_{E}=\bigl\langle\Delta r_{y}(t)^{2}\bigr\rangle_{E}$.
In Figure \ref{fig:NEMD_msd}, we plot $\bigl\langle\Delta r_{\parallel}(t)^{2}\bigr\rangle_{E}$
and $\bigl\langle\Delta r_{\perp}(t)^{2}\bigr\rangle_{E}$ as a function
of time for an example system at various electric field strengths
$E=0.05$--$0.20$, which at sufficiently long times reach log-log
slopes of 2 and 1, respectively. To obtain drift velocity
  $\left\langle v_{x}\right\rangle$, the long time logarithmically
  spaced data is fitted to $a+bt^{2}$, where $a$
  and $b$ are fitting parameters and $\left\langle v_{x}\right\rangle=\sqrt{b}$. 
The average ion mobility $\mu$ can then be obtained by
\begin{equation}
\mu=\frac{\bigl\langle v_{x}\bigr\rangle}{E}
\end{equation}
We are interested in $\mu$ in the linear response regime, in which the
system's behavior is not significantly disrupted from its equilibrium
behavior, other than the ions' drift in response to the field. Thus, it is necessary to
check whether the drift velocity is linear with respect to $E$; in
this case,
value of $\mu$ does not depend on $E$. 
We found that our systems are well in the linear response regime across
a wide range of ion concentrations and Coulomb strengths, as shown
in the Supporting Information.

The molar conductivity can be directly
calculated from mobilities using

\begin{equation}
\Lambda=\frac{\lambda}{N_{ion}}=\frac{e(\mu_{+}+\mu_{-})}{V}\label{eq:cond_NEMD}
\end{equation}
where $\mu_{+}$ and $\mu_{-}$ are cation and anion mobilities, respectively
($\mu_{+}=\mu_{-}=\mu$ in this work). Then, $\Lambda/\Lambda_{NE}$
in NEMD is the ratio between eq \ref{eq:cond_NEMD} and eq \ref{eq:cond_NE},
equivalent to
\begin{equation}
\left(\frac{\Lambda}{\Lambda_{NE}}\right)_{NEMD}=\frac{k_{B}T}{e}\left(\frac{\mu_{+}+\mu_{-}}{D_{+}+D_{-}}\right)\label{eq:alpha_NEMD}
\end{equation}
where the diffusion constants, $D_{+}$ and $D_{-}$, are calculated
from eq \ref{eq:diffusion_Einstein} using y- and z-directional MSD
for HP and y-directional MSD for BCP. We note that the MSDs used to
estimate diffusion constant have one degree of freedom less than those
in EMD. 

\subsection{Ion Cluster and Polymer Relaxation Rate }

To show the ion cluster and polymer dynamics, we calculated the cluster
autocorrelation function, $ACF_{cluster}$ and monomers' self-intermediate
scattering function, $S(k,t)$. We define $ACF_{cluster}=\sum_{i\neq j}C_{ij}(t)$,
where $C_{ij}(t)=1$ if the beads $i$ and $j$ are in the same cluster
at both time $t$ and at time $0$, and $C_{ij}(t)=0$ otherwise;
the sum is over all ions and normalized to be 1 at time 0. In line
with prior work, ions were considered in the same cluster if they
were within $1.05\sigma$ from any other ions in the cluster, and
the initial data before $0.15\tau$ was removed for better fitting
(neglecting the rapid decorrelation at very short times).

We also calculated the self-intermediate scattering function by
\begin{equation}
S(k,t)=\frac{1}{N_{mon}}\sum_{i=1}^{N_{mon}}\bigl\langle e^{-i\mathbf{k}\cdotp\Delta r_{i}(t)}\bigr\rangle
\end{equation}
where $N_{mon}$ is the number of monomer beads, $\mathbf{k}$ is
the scattering vector with a magnitude of $k$. We set $k=\pi$ so
that $S(k,t)$ decayed slowly enough for good fitting. Both $ACF_{cluster}$
and $S(k,t)$ were fitted to a scaled Kohlrausch-Williams-Watts (KWW)
stretched exponential function, $\alpha exp[-(t/\tau^{*})^{\beta}]$,
where $\alpha$, $\beta$, and $\tau^{*}$ are the fitting parameters.
The ion cluster relaxation rate $\tau_{cluster}^{-1}$ or polymer relaxation rate $\tau_{polymer}^{-1}$ was then calculated
by $\tau_{cluster/polymer}^{-1}=[(\tau^{*}/\beta)\Gamma(1/\beta)]^{-1}$, where
$\Gamma(x)$ is the gamma function.

\section{Results and Discussion}

\subsection{Equilibrium vs Nonequilibrium Method}

\begin{figure}[H]
\includegraphics{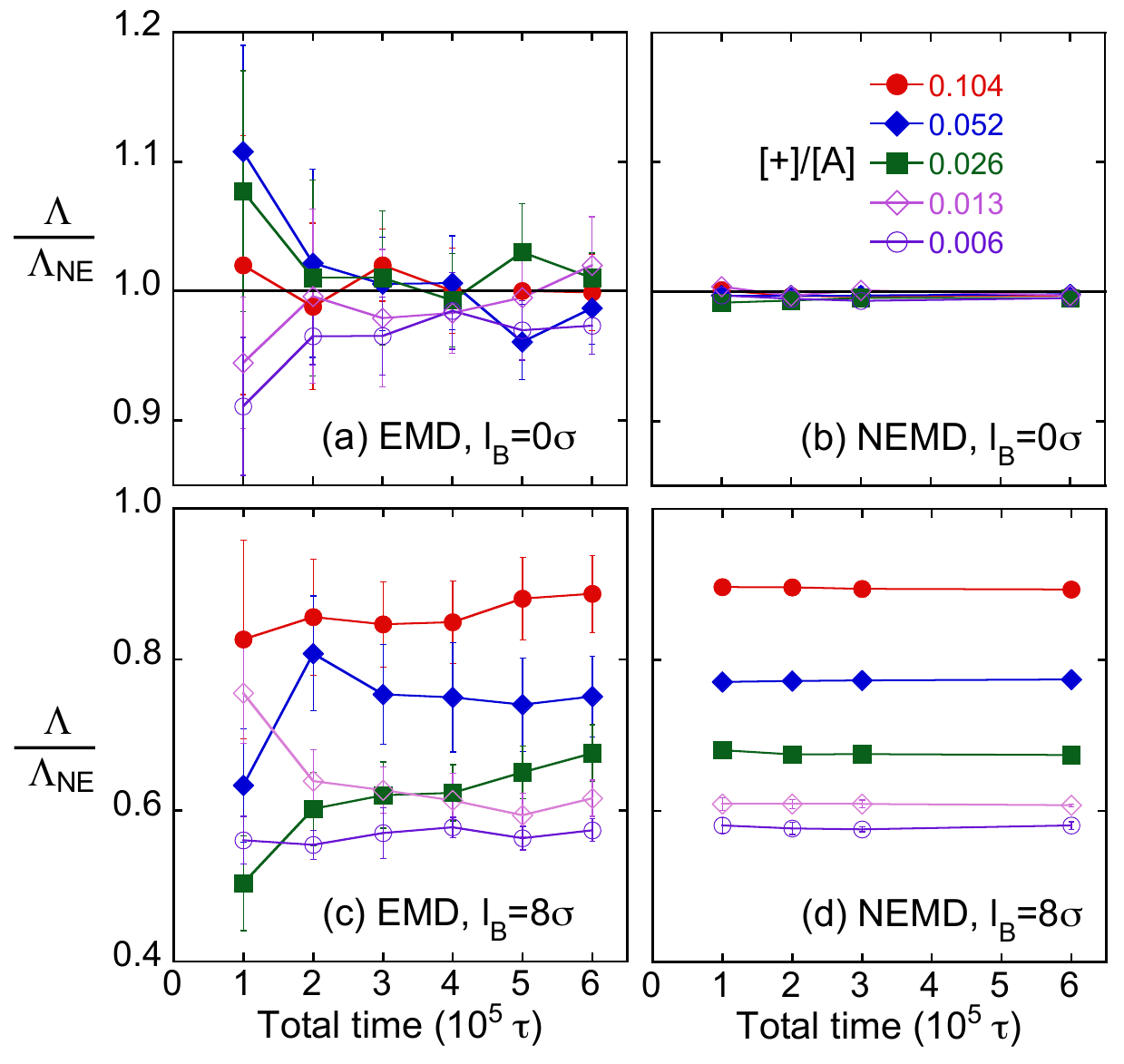}

\caption{The degree of uncorrelated ion motion (fractional deviation of conductivity
from the Nernst-Einstein equation) $\Lambda/\Lambda_{NE}$ as a function
of total time of data collection (excluding the equilibration
time) in homopolymers at various ion concentrations for (a),(b) $l_{B}=0\sigma$
and (c),(d) $l_{B}=8\sigma$ from EMD (a),(c) and NEMD (b),(d). The
error bars are the standard errors across 8 systems with different
initial configurations.\label{fig:method_comparison}}
\end{figure}

As a first step to assess the accuracy and reliability of conductivity
calculation from both equilibrium and nonequilibrium methods, we calculate
$\Lambda/\Lambda_{NE}$ for toy systems with no Coulomb interactions
between ions ($l_{B}=0\sigma$). Under such condition, ion mobility
is related to the ion self-diffusion constant via the Einstein relation:
$D=\mu k_{B}T/e$, which grants $\Lambda/\Lambda_{NE}=1$ regardless
of salt loading. Thus, checking
how close to 1 $\Lambda/\Lambda_{NE}$ is allows us to assess accuracy. In Figure \ref{fig:method_comparison}a
and b, we show $\Lambda/\Lambda_{NE}$ at $l_{B}=0\sigma$ calculated
using different total amounts of simulation time in EMD and NEMD, respectively.
More blocks (time windows) are averaged over for the $\Lambda/\Lambda_{NE}$
calculation as total time increases, and the error bars are the standard
errors across 8 systems with different initial configurations. For
EMD, the mean of $\Lambda/\Lambda_{NE}$ ranges between 0.9 and 1.1
with errors of $\pm5$--$10\%$ when calculated using the shortest simulation time considered. The errors are $\pm10$--$20\%$
when considering 95\% confidence intervals, which are close to those
reported in Ref \citenum{france2019correlations}. Although the errors
reduce and the means converge closer to 1 as more data is averaged over, there
is a persistent error even at the longest times considered here. However,
$\Lambda/\Lambda_{NE}$ from NEMD is significantly faster
to converge to 1 with small error compared to that from EMD.
We note that, due to the concurrent calculation of $D$ and $\mu$ by assessing motion perpendicular and parallel to the field, the NEMD and EMD simulations are run for the same total time. Applying the field has negligible effect on the calculation time. The calculation of $\mu$ as a function of $E$ to ensure the system is in the linear response regime, however, is an additional step required for NEMD. This assessment did not take very long compared to the overall simulation time (see the Supporting Information). In short, the toy model results show that calculating conductivity from nonequilibrium
method is rather accurate and reliable.

We then computed $\Lambda/\Lambda_{NE}$ at $l_{B}=8\sigma$ using
both methods, as shown in Figure \ref{fig:method_comparison}c and
d. Unlike the toy systems where $\Lambda/\Lambda_{NE}=1$
at all ion concentrations, the result for $l_{B}=8\sigma$ is not known a priori. For EMD, the large error bars at various concentrations overlap,
leading to the difference in some systems' $\Lambda/\Lambda_{NE}$
statistically insignificant. In contrast, $\Lambda/\Lambda_{NE}$
from NEMD is consistent after $10^{5}\tau$ of simulation time with negligible error on the scale of the plot. This again shows that it is much more efficient to calculate conductivity from
NEMD (it takes about 6 times longer for EMD to reach similar mean
values but still with relatively large errors).

\subsection{Ion Transport Properties from NEMD}

After validating the conductivity calculation on a set of systems above, we now use the NEMD method to analyze ion transport in a variety of salt-doped polymer systems;
specifically, we probe the effects of Coulomb and solvation interactions
on ion transport by separately adjusting $l_{B}$, $S_{A\pm}$,
and $S_{\pm\pm}$. We note that the mapping approach discussed in the Simulation Model section implies that $l_{B}$ and $S_{A\pm}$ are not independent (both depend on the dielectric constant of the A phase), but we vary them separately here to show the effects of these parameters. All results from the following sections are from
NEMD simulations with ion conductivity directly calculated from ion
mobility.

\subsubsection{Effects of Coulomb strength}

\begin{figure}[H]
\includegraphics{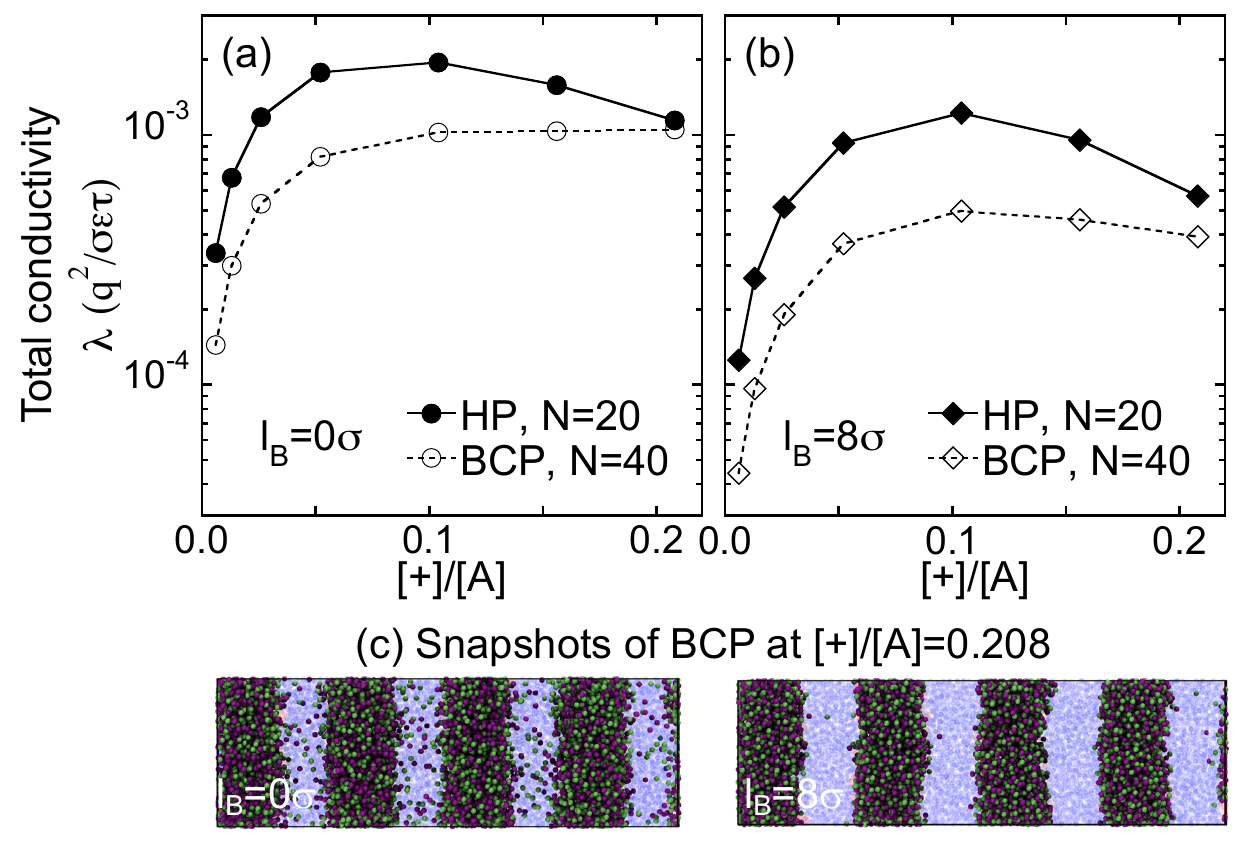}

\caption{Ion conductivity in homopolymers (solid symbols) and block copolymers
(open symbols) across a range of ion concentrations at (a) $l_{B}=0\sigma$
and (b) $l_{B}=8\sigma$. (c) Snapshots of the block copolymer systems
at $[+]/[A]=0.208$.\label{fig:cond_bcpvshp}}
\end{figure}

\begin{figure}[H]
\includegraphics{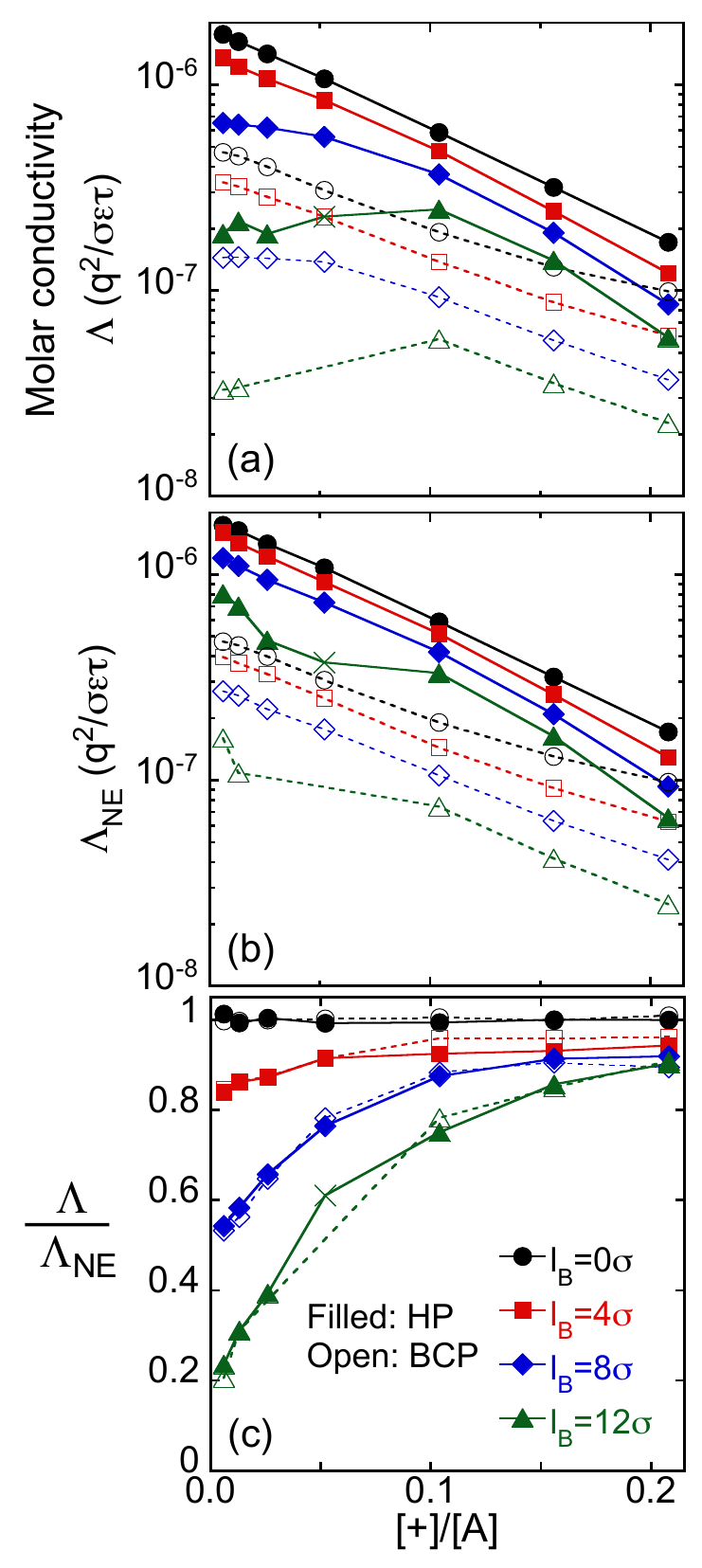}

\caption{(a) Molar conductivity $\Lambda$, (b) Nernest-Einstein molar conductivity
$\Lambda_{NE}$, and (c) degree of uncorrelated ion motion $\Lambda/\Lambda_{NE}$
in homopolymers (filled symbols) and in block copolymers (open symbols)
as a function of ion concentration at $l_{B}=0\sigma$ (circles),
$l_{B}=4\sigma$ (squares), $l_{B}=8\sigma$ (diamonds), and $l_{B}=12\sigma$
(triangles).\label{fig:molarcond_alllB}}
\end{figure}

Figure \ref{fig:cond_bcpvshp}a and b shows the total ion conductivity
in both HP and BCP systems across a wide range of salt loading at
$l_{B}=0\sigma$ and $l_{B}=8\sigma$, respectively. Both HP curves reproduce the experimentally observed nonmonotonic
trend, and
we find the total conductivity $\lambda$ in HPs at both Coulomb strengths
peaks at a similar salt loading. The nonmonotonic relationship between
conductivity and concentration apparently stems from the competition between
the decrease in ion diffusion with concentration and the increase in number of ions.
Prior all-atom simulations also have shown nonmonotonic conductivity
behavior with relatively high $\Lambda/\Lambda_{NE}$ ($\approx0.8$--$1.0$).\cite{borodin2006mechanism}
However, in BCPs, the $\lambda$ trend varies with Coulomb strength. At
$l_{B}=0\sigma$, $\lambda$ plateaus at high concentrations,
whereas at $l_{B}=8\sigma$, $\lambda$ is nonmonotonic, which
matches experiment.\cite{chintapalli2016structure,galluzzo2020measurement} We note that the
conductivity of BCPs is typically smaller than that of HPs, as
expected from prior work.\cite{chintapalli2016structure,singh2007effect} The BCP conductivity becomes similar
to that of HPs at high concentration
at $l_{B}=0\sigma$, which might due to ions partly mixing in the nonconducting
B phase at high concentration, as shown in Figure \ref{fig:cond_bcpvshp}c.
This amount of mixing observed in our toy model at $l_{B}=0\sigma$ is
not expected to be present in the BCP electrolyte systems of 
interest, which include both Coulomb interactions and strong selective
solvation interactions.\cite{gomez2009effect}
To better understand how Coulomb strength affects ion correlations,
we plotted molar conductivity $\Lambda$ and Nernst-Einstein molar
conductivity $\Lambda_{NE}$ at different $l_{B}$ values in both
homopolymer and block copolymer systems in Figure \ref{fig:molarcond_alllB}a
and b, respectively (data of ion mobility and diffusion constant used
to calculate conductivity are available in the Supporting Information).
We find linear $\Lambda$ trends on a log scale at weak Coulomb strengths
and nonlinear trends at strong Coulomb strengths. On the other hand,
$\Lambda_{NE}$ as a function of ion concentration is always linear
(except for the intermediate concentration regime at $l_{B}=12\sigma$,
which can be attributed to the formation of larger ion clusters and
will later be discussed). Linear $\Lambda_{NE}$ behavior is somewhat
expected as $\Lambda_{NE}$ is mainly determined by the ion diffusion
constant, which is known to be linear versus ion concentration on
a log scale.\cite{borodin2006mechanism} A cross symbol is used to
present an approximate result for the system at $l_{B}=12\sigma$ and $[+]/[A]=0.052$. Around
this intermediate salt loading and strong Coulomb strength, the system
appears to be macrophase separating. Specifically, large ion-rich and
ion-poor regions are observed, and depending on whether the ion-rich
region is connected from one side of the box to the other, the
conductivity measurement can vary. Without assessing different box
sizes to understand the overall
structure of a macroscopic system, which we have not done, we cannot
be confident in the bulk conductivity for this system. 
For the approximate result shown, we used the average from two samples
where the ion-rich region is connected through the box in the parallel and perpendicular
directions to the electric field, respectively (see the Supporting Information
for more detail). The conductivity data from this system is consistently represented by the cross symbol throughout the work. However, in terms of the structural results, cluster relaxation rate, and polymer relaxation rate, the data do not depend significantly on the overall morphology of the ion-rich region. Thus, we have included this system as a data point in our other results.

Figure \ref{fig:molarcond_alllB}c summarizes the degree of uncorrelated
ion motion for each system,
the ratio $\Lambda/\Lambda_{NE}$. As expected, increasing Coulomb strength increases ion correlations (decreases $\Lambda/\Lambda_{NE}$). Interestingly, we find that $\Lambda/\Lambda_{NE}$ is generally lower at low ion
concentrations (except for $l_{B}=0\sigma$ where it has to
be $1.0$ regardless of salt loading), indicating that ion motion
is more correlated at low concentrations. This contradicts the general
understanding in typical electrolytes that ion motion is less correlated
at low concentrations due to the screening of solvent, and is more
correlated at high concentrations as large ion clusters can form.\cite{france2019correlations}
While it is theoretically true that ion motion should become uncorrelated in the dilute limit as ions are spread far apart, we speculate that ion concentration would have to be extremely low for ion
correlations to be negligible in these salt-doped polymers with strong
ion-polymer and ion-ion interactions. Nonetheless, a similar trend of ion correlations has
also been reported in studies of ionic liquid electrolytes\cite{haskins2014computational}
and nonaqueous polyelectrolytes.\cite{fong2019ion} In the latter
work, the contribution to conductivity from cation-anion correlated
motion is more negative at lower ion concentration.\cite{fong2019ion}
To probe the underpinnings of these results, we analyze ion
cluster and polymer relaxation rate. we found little difference in these relaxation times between HP and BCP systems, and we show only the HP results for the following
sections (BCP results are shown in the Supporting Information).

\begin{figure}[H]
\includegraphics{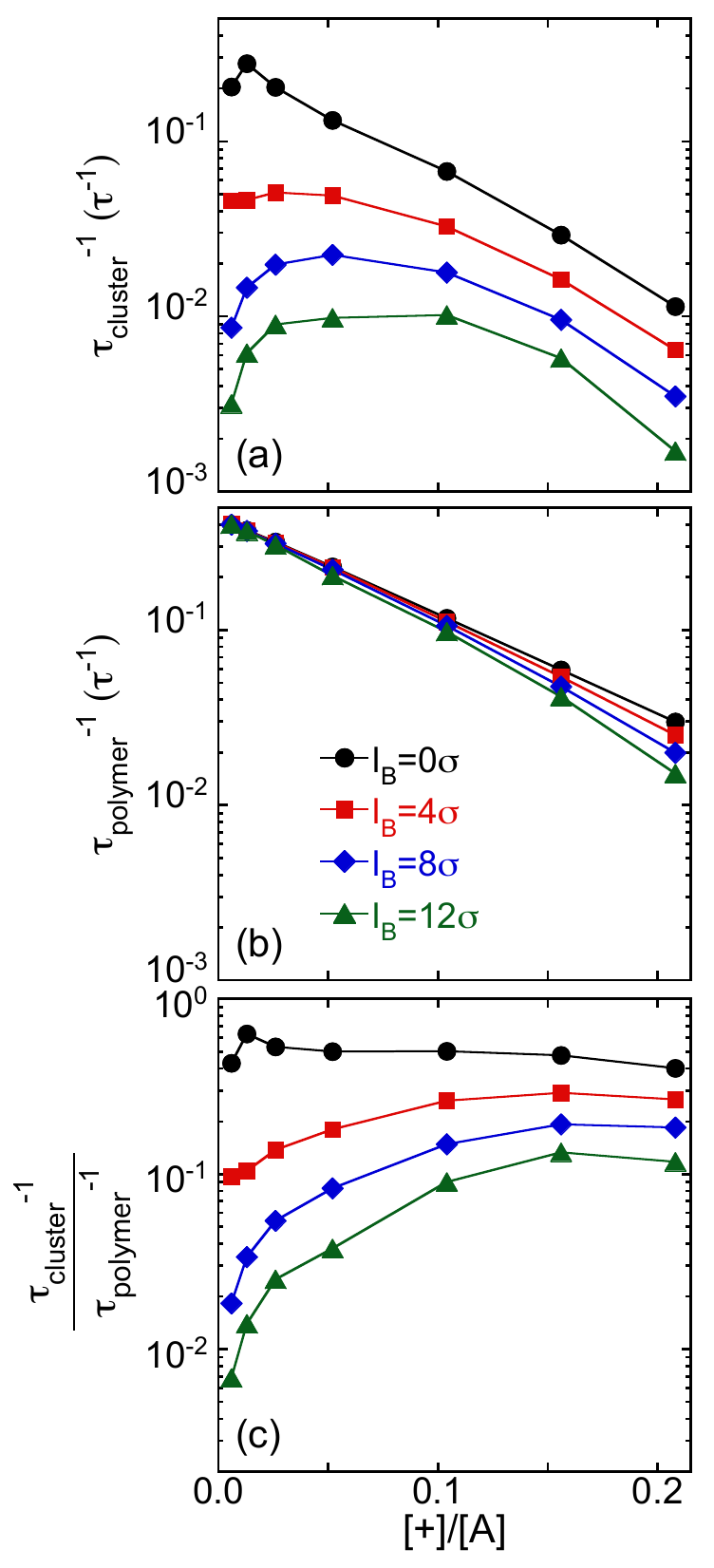}

\caption{(a) Ion cluster relaxation rate, (b) polymer relaxation rate, and (c) their ratio (normalized ion cluster relaxation rate) as
a function of ion concentration at different Coulomb strengths. \label{fig:relaxation_rates}}
\end{figure}

To separately study ion and polymer dynamics, we calculated cluster
relaxation rate $\tau_{cluster}^{-1}$ and polymer relaxation rate
$\tau_{polymer}^{-1}$ from the ion cluster autocorrelation function and
monomers' self-intermediate scattering function. These results are shown in Figure
\ref{fig:relaxation_rates}. We find that increasing
salt loading or Coulomb strength generally lowers $\tau_{cluster}^{-1}$.
However, $\tau_{cluster}^{-1}$ also decreases in the lower concentration
regime at strong Coulomb strength, leading to a nonmonotonic trend.
As detailed in the Supporting Information, we obtained similar nonmonotonic
results using different definitions of ion cluster relaxation rates
that have been reported to be related to ion conductivity.\cite{zhao2009there,zhang2015direct}
On the other hand, the behavior of $\tau_{polymer}^{-1}$ with concentration
is linear on a semi-log scale and is almost independent of Coulomb strength, signifying
polymer dynamics mainly depend on salt loading rather than ion dynamics
(until very high concentration). We also find that $\tau_{polymer}^{-1}$
is proportional to ion diffusion (in the Supporting Information)
or $\Lambda_{NE}$ (Figure \ref{fig:molarcond_alllB}b) across various
systems. This is similar to recent experimental results showing
that, even in samples with inhomogeneous salt distributions, the product
of lithium ion diffusion and polymer relaxation time is temperature-independent
in lithium salt-doped poly(propylene glycol).\cite{becher2019local}

We then normalized cluster relaxation rate by polymer relaxation rate,
as shown in Figure \ref{fig:relaxation_rates}c. The normalized cluster
relaxation rate versus concentration is flat at $l_{B}=0\sigma$,
indicating that ion dynamics is highly dependent on polymer dynamics
for ions with no electrostatic interactions. However, the normalized
rate reduces with increasing Coulomb strength and decreasing ion concentration
for other systems. Interestingly, we find the behavior of the normalized
cluster relaxation rate highly resembles that of $\Lambda/\Lambda_{NE}$,
shown in Figure \ref{fig:molarcond_alllB}c. The result suggests that
the correlation in ion motion is closely related to how long ion clusters
remain intact, and is in line result from Ref. \citenum{fong2019ion}
where correlated cation-anion pairs were found to be able
to travel longer distances with decreasing concentration despite the
reduced fraction of ion pairs. The relationship between $\Lambda/\Lambda_{NE}$
and cluster relaxation rate will further be discussed in the following
sections.

\begin{figure}[H]
\includegraphics[scale=0.75]{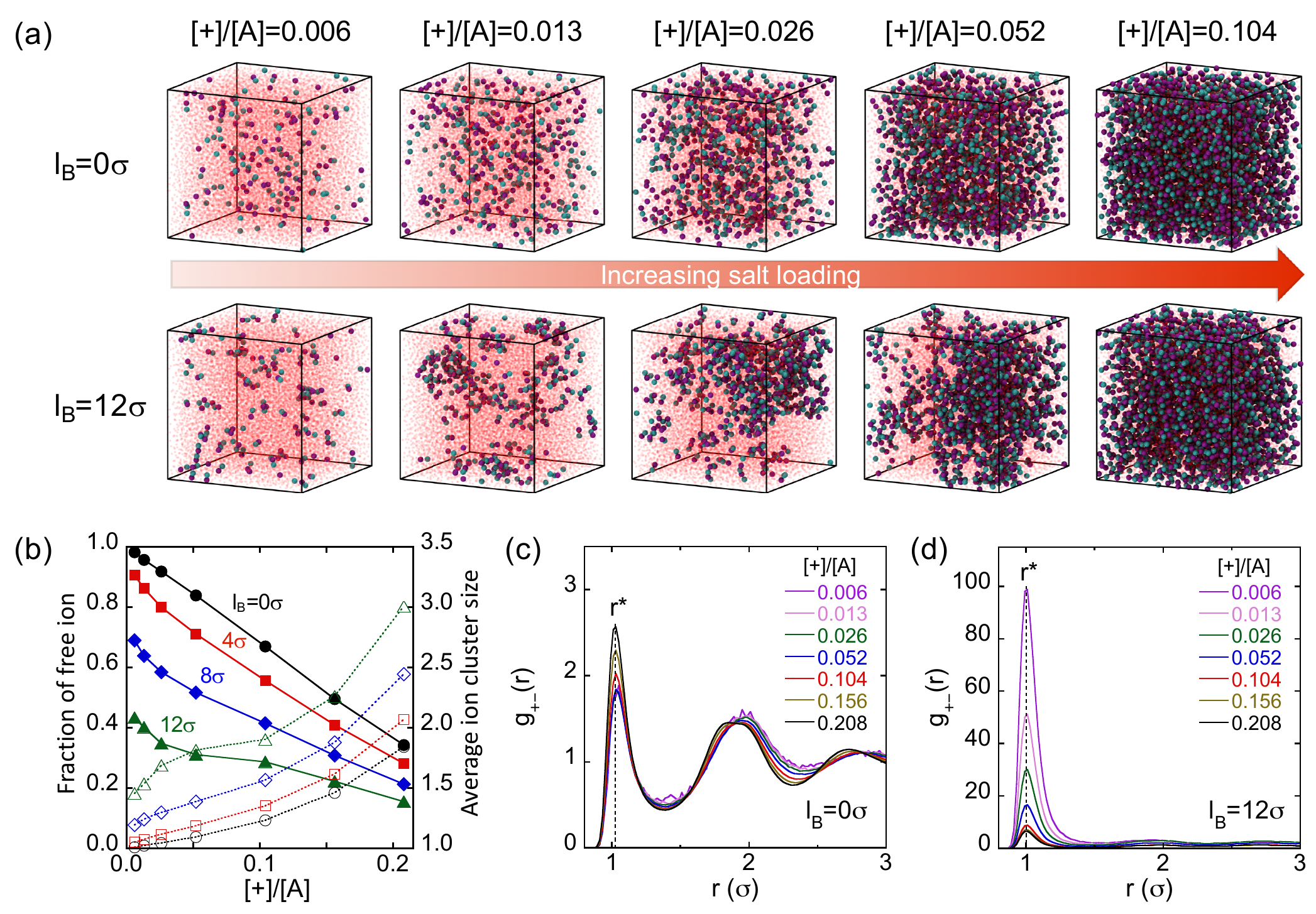}

\caption{(a) Snapshots of homopolymer systems with various salt loading at
$l_{B}=0\sigma$ (top) and $l_{B}=12\sigma$ (bottom). (b) Fraction
of free ions (solid lines) and average ion cluster size (dashed lines)
versus salt loading for $l_{B}=0$--$12\sigma$. (c), (d) Cation-anion radial distribution functions of
homopolymer systems with various salt loading at $l_{B}=0\sigma$
and $l_{B}=12\sigma$, respectively.\label{fig:structural_properties}}
\end{figure}

The cation-anion motion correlation behavior with concentration shown
in Figure \ref{fig:molarcond_alllB}c is somewhat nonintuitive as one
would expect that the fraction of free ions would decrease (or the cluster
size to increase) as salt loading increases and that this would increase
correlation in ion motion. To fully understand the trend, we analyzed a variety
of structural properties of systems with different Coulomb strengths,
as shown in Figure \ref{fig:structural_properties}. Indeed, we find
the fraction of free ions decreases and the average ion cluster size
increases with growing ion concentration regardless of Coulomb strength
(Figure \ref{fig:structural_properties}b). Meanwhile, stronger Coulomb
strength consistently makes ions more clustered at all concentrations. 

\begin{figure}[H]
\includegraphics{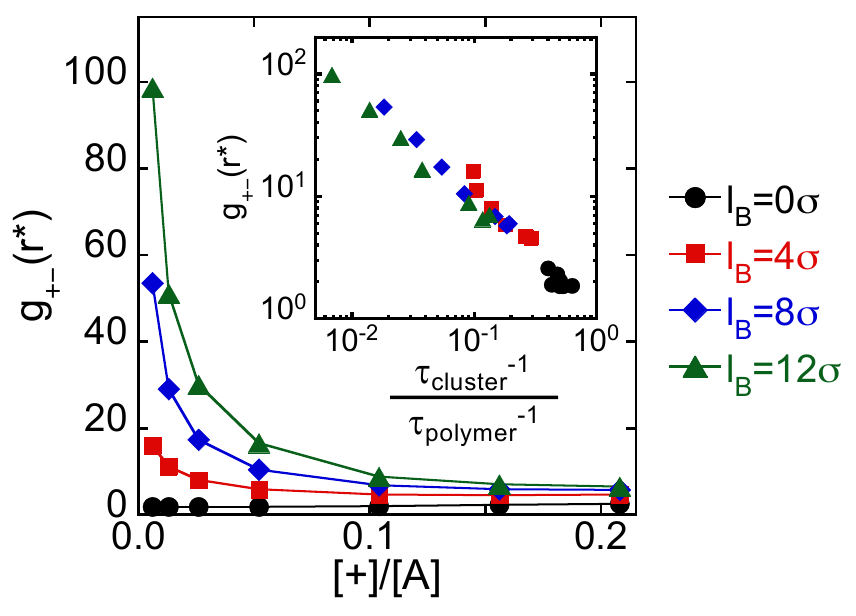}

\caption{The first peak height of the cation-anion radial distribution function versus ion concentration at $l_{B}=0$, $4$, $8$,
and $12\sigma$ with the inset showing its log-log scale relationship
with the normalized cluster relaxation time.\label{fig:g(r)_peakheight}}
\end{figure}

We also consider ion structure by comparing the cation-anion radial distribution function, $g_{+-}(r)$,
at different salt loadings and Coulomb strengths in Figure \ref{fig:structural_properties}, along with snapshots.
At $l_{B}=0\sigma$ (Figure \ref{fig:structural_properties}c), systems
at various salt loadings share a similar first peak height, denoted
as $g_{+-}(r^{*})$. Only at very high concentration does $g_{+-}(r^{*})$
slightly increase, potentially due to the relative saturation of available
sites for ion-monomer contact leading to some ion-ion contact.
On the other hand, at $l_{B}=12\sigma$ (Figure \ref{fig:structural_properties}d),
$g_{+-}(r^{*})$ varies significantly with salt loading; specifically, the first peak height is higher at low concentrations. Figure \ref{fig:g(r)_peakheight} shows the height of the first peak in $g_{+-}(r)$ at different Coulomb strengths. Increasing Coulomb strength
significantly increases the peak in the low concentration
regime, and then the peak height plateaus after $[+]/[A]=0.1$ for all
systems. Interestingly, we find the peak height, a structural measure of the amount of ion contacts, is related to cluster dynamics, specifically to the normalized
cluster relaxation rate. In fact, all data points can be approximately collapsed
on a single curve (inset figure) when plotting $g_{+-}(r^{*})$ versus normalized cluster relaxation rate.

\begin{figure}[H]
\includegraphics{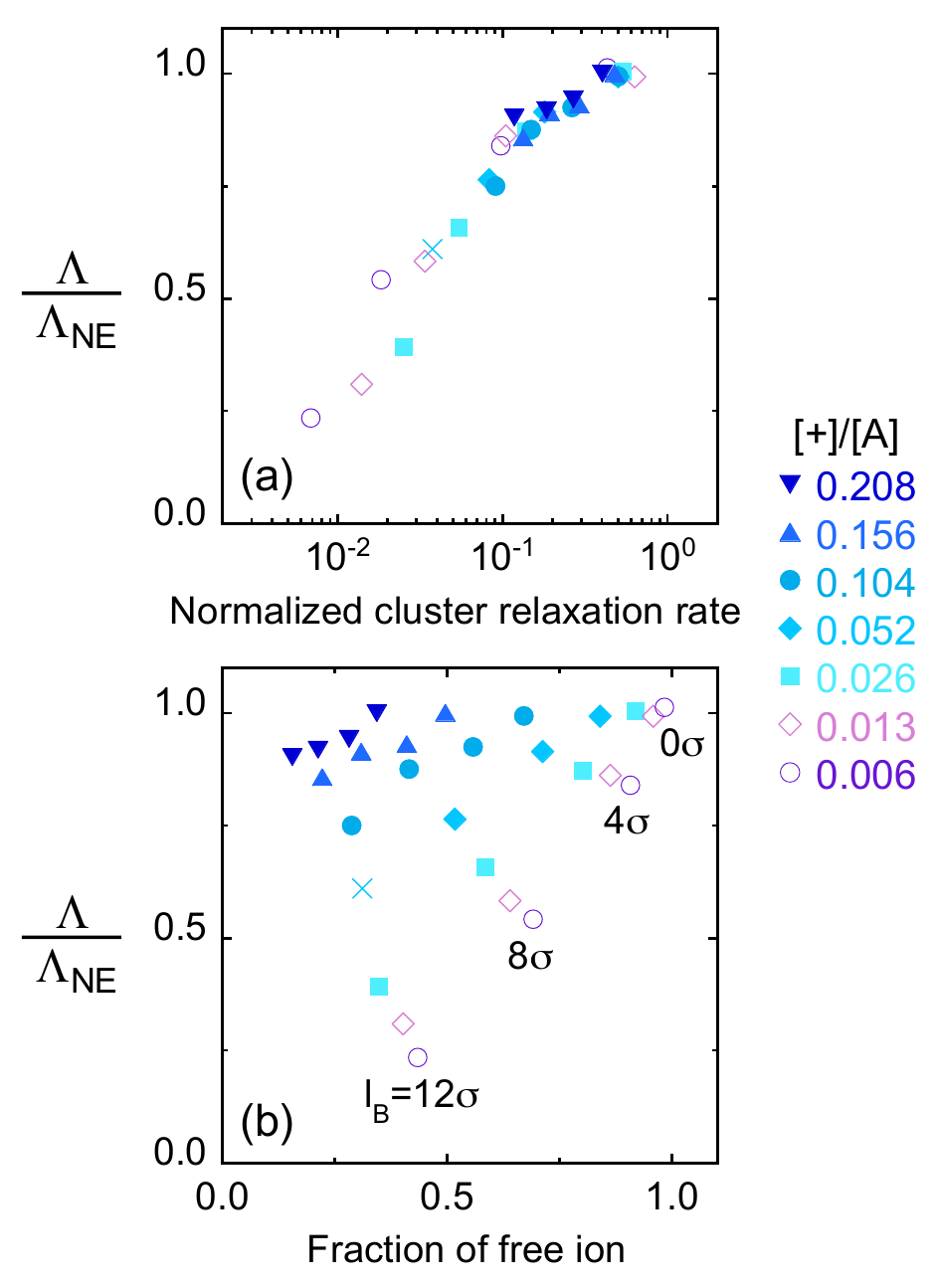}

\caption{The degree of uncorrelated ion motion $\Lambda/\Lambda_{NE}$ as a
function of (a) normalized cluster relaxation rate and (b)
fraction of free ions across a range of Coulomb strengths and
ion concentrations. Going from right to left for each symbol, $l_{B}=0$, $4$, $8$,
then $12\sigma$. \label{fig:alpha_connection}}
\end{figure}

To further understand how ion motion correlation connects to other
properties, we first plotted $\Lambda/\Lambda_{NE}$ versus normalized
cluster relaxation rate, as shown in Figure \ref{fig:alpha_connection}a. A strong correlation was found between the two properties across
all systems regardless of Coulomb strength and ion concentration,
indicating correlated ion motion is closely related to cluster dynamics.  Because normalized cluster relaxation is also related to the height of the first peak of $g_{+-}(r)$, this also means that $g_{+-}(r^{*})$ predicts $\Lambda/\Lambda_{NE}$. However, the normalized cluster relaxation better explains the $\Lambda/\Lambda_{NE}$ data, especially for BCP systems whose $g_{+-}(r^{*})$ is apperently affected by ion structure near interfaces (see the Supporting Information).

One may also expect conductivity to be related to the free ion content or ion dissociation. As a result, we have also
plotted $\Lambda/\Lambda_{NE}$ against fraction of free ions at different
Coulomb strengths and ion concentrations. Surprisingly, there is a negative correlation
between $\Lambda/\Lambda_{NE}$ and fraction of free ions as a function of ion concentration
at each Coulomb strength above 0 (Figure
\ref{fig:alpha_connection}b). These
results show that ion motion can be less correlated when the fraction of free
ions is lower (corresponding to higher concentrations and ions being more clustered), which is counterintuitive
as ion agglomeration is often considered to deteriorate the ion transport.
Apparently, the strength of the local ion association as quantified by $g_{+-}(r^{*})$ or normalized cluster relaxation time is a better predictor of conductivity than the fraction of free ions.

\subsubsection{Effects of ion-monomer and ion-ion solvation strength}

\begin{figure}[H]
\includegraphics{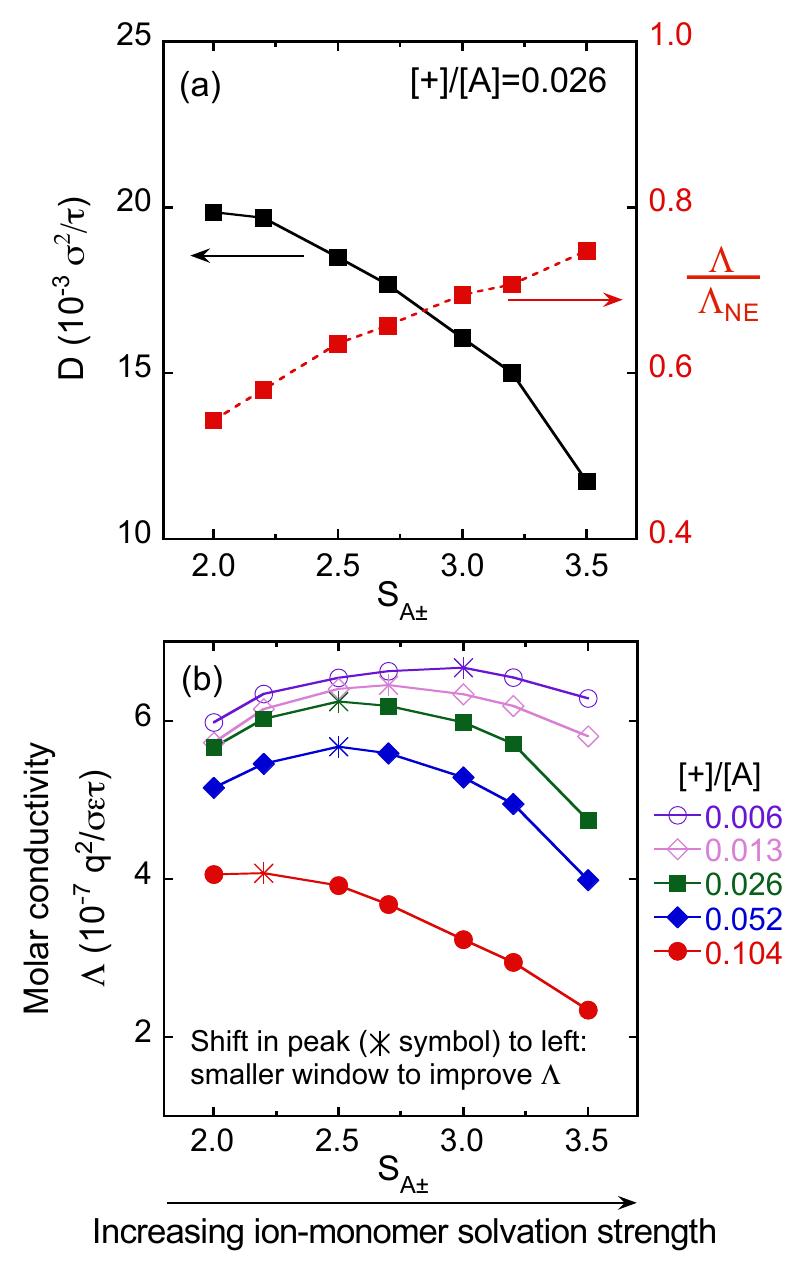}

\caption{(a) Diffusion constant (solid line, left axis) and degree of uncorrelated
ion motion (dashed line, right axis) at $[+]/[A]=0.026$ and (b) molar
conductivity at various ion concentrations, as a function of ion-monomer solvation strength in homopolymers.\label{fig:Sion_monomer_effect}}
\end{figure}

We now probe the effects of ion-monomer solvation interaction on ion
conductivity by adjusting ion-monomer solvation strength $S_{A\pm}$
and holding other variables constant. As shown in Figure \ref{fig:Sion_monomer_effect}a,
ion diffusion constant decreases and $\Lambda/\Lambda_{NE}$ increases
with increasing $S_{A\pm}$. In line with the results of Wheatle, Lynd, and Ganesan in Ref. \citenum{wheatle2018effect}, we also find that
stronger ion-monomer interaction reduces correlated ion motion but
slows polymer dynamics as well as ion diffusion. Figure \ref{fig:Sion_monomer_effect}b
shows molar conductivity $\Lambda$ as a function of $S_{A\pm}$ at
different ion concentrations. Increasing ion concentration increases $\Lambda/\Lambda_{NE}$ but decreases diffusion
constant and $\Lambda_{NE}$. Overall, $\Lambda$ tends to peak at intermediate ion-monomer
solvation strength. Significantly, we find the peak shifts to lower
$S_{A\pm}$ with increasing ion concentration. As detailed in Supporting
Information, at higher concentration, the detrimental effect of retarded
ion diffusion becomes more influential than the conducitvity-promoting
effect from reduced ion correlations. The peak shifting suggests
that the potential materials design strategy of improving ion conduction by using polymers with higher dielectric constant (strengthening ion-monomer
interaction) will have limited effectiveness at
high salt loading. Instead, ion conductivity at high concentrations is mainly
limited by the diffusion constant for the systems considered here.

We also show the effects of ion-ion solvation strength $S_{\pm\pm}$
on ion transport in the Supporting Information. Nonzero $S_{\pm\pm}$ can be chosen because some ions may be bulky, polarizable molecules and thus ``solvate'' other ions in addition to interacting with them through the Coulomb potential. As intuitively expected, increasing
$S_{\pm\pm}$ not only slows ion diffusion but also intensifies correlations in ion motion (see the Supporting Information). Thus, the molar conductivity decreases monotonically
with larger $S_{\pm\pm}$. However, we postulate that this might not be the case if there were a significant size disparity between cation and anion. With small cations, their strong complexation with monomers may be the main factor acting to impede their diffusion, and we hypothesize it may be possible to mitigate this effect by including stronger interactions with anions, leading to a situation in which increasing cation-anion interactions can increase diffusion in a particular range of parameter space. We hope to explore these possible effects in future work.

\begin{figure}[H]
\includegraphics[scale=0.75]{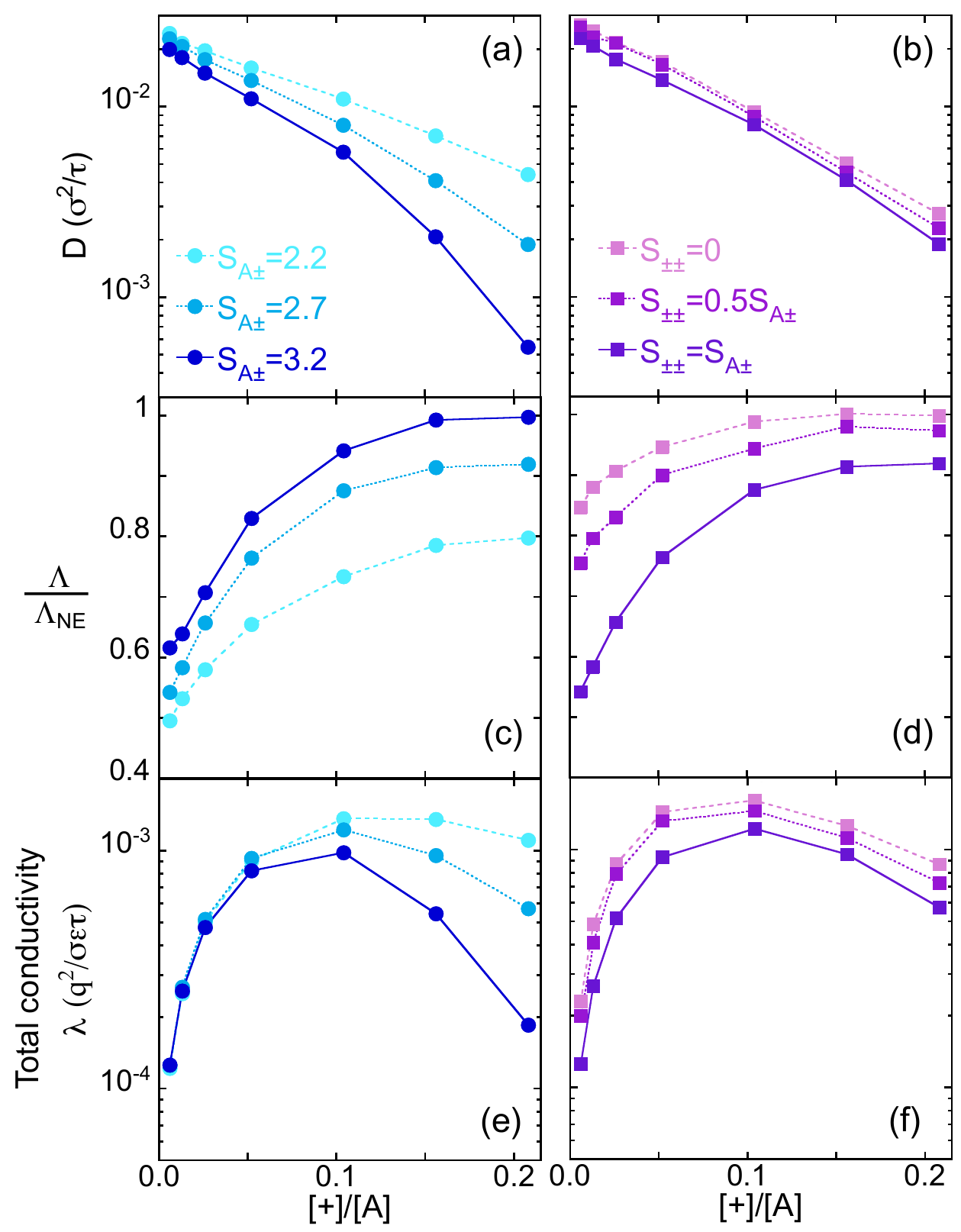}

\caption{(a),(b) Diffusion constant, (c),(d) degree of uncorrelated ion motion
$\Lambda/\Lambda_{NE}$, and (e),(f) ion conductivity as a function
of ion concentration at various ion-A interactions (a),(c),(e) and
ion-ion interactions (b),(d),(f).\label{fig:totalcond_vs_solvation}}
\end{figure}

The total conductivity $\lambda$ and other transport properties
were measured for several selected systems with different $S_{A\pm}$
and $S_{\pm\pm}$ values across a wide range of ion concentration,
as shown in Figure \ref{fig:totalcond_vs_solvation}. We find that tuning
ion-monomer or ion-ion interactions leads to different total conductivity
versus ion concentration results: 1) Increasing $S_{A\pm}$ reduces
ion motion correlation as discussed earlier, yet slows ion diffusion,
especially at high concentrations. The resulting $\lambda$ remains
similar with stronger ion-monomer interactions in the lower concentration
regime due to competing effects between reduced diffusion constant
and more uncorrelated ion motion. However, $\lambda$ decreases in the higher concentration regime. 2) Increasing $S_{\pm\pm}$
causes less uncorrelated cation-anion motion as well as slower ion
diffusion; $\lambda$ shifts down to a similar extent at all
concentrations.

We note that we expect the trend in HPs shown here to be similar as
that in BCPs based on the results in Figure \ref{fig:molarcond_alllB}
that both ion diffusion and $\Lambda/\Lambda_{NE}$ trends are similar
for HP and BCP systems (except the diffusion constant in BCP is always
smaller). In addition, from our previous
studies,\cite{brown2018ion,seo2019role} we have shown that the dielectric
contrast $\Delta S$ (related to the difference in dielectric constant
between the two blocks in BCP) is the key factor that determines how ion
transport properties in BCPs deviate from those in HPs. Thus, we also have tested
the impact of $\Delta S$ on ion dynamics in this work, as shown in
the Supporting Information. In line with Ref. \citenum{seo2019role},
we show that $\Lambda/\Lambda_{NE}$ is nearly independent of $\Delta S$
and ion diffusion constant decreases with increasing $\Delta S$.

\subsubsection{Ion Correlations in Low and High Concentration Regimes}

\begin{figure}[H]
\includegraphics{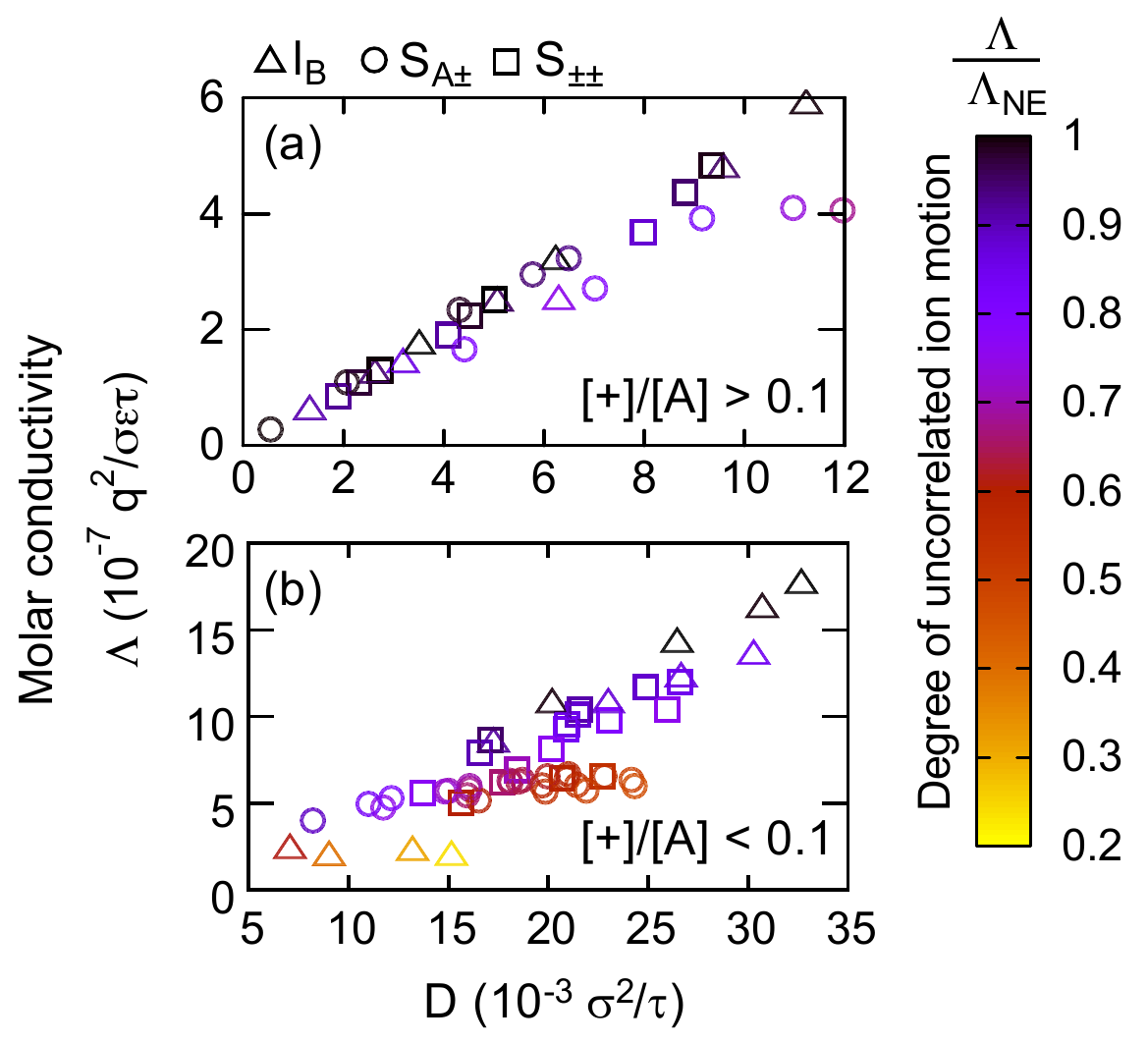}

\caption{Molar conductivity as a function of ion diffusion constant at (a)
high and (b) low concentration. The color of the data points
corresponds to the degree of uncorrelated ion motion as indicated
by the color bar. Points with different shapes represent systems with
varying Coulomb strength (triangles), ion-monomer solvation strength
(circles), and ion-ion solvation strength (squares).\label{fig:overall_alpha}}
\end{figure}

Finally, having observed the dependence of ion correlations on salt
loading, we plotted molar conductivity of all systems considered above
(including systems with various $l_{B}$, $S_{A\pm}$, and $S_{\pm\pm}$)
as a function of ion diffusion constant separately at high concentrations
($[+]/[A]>0.1$) and at low concentrations ($[+]/[A]<0.1$), as shown
in Figure \ref{fig:overall_alpha}a and b, respectively. At high concentrations,
regardless of changes in interactions, we find molar conductivity
can be well predicted by diffusion constant since $\Lambda/\Lambda_{NE}$
is close to 1. On the other hand, at low concentrations, $\Lambda/\Lambda_{NE}$
ranges from $0.2-1.0$ and molar
conductivity is not always well predicted by the diffusion constant. Out another way, this means that the Nernst-Einstein
equation tends to be a better approximation at high concentrations in polymer electrolytes,
whereas at low concentrations, ion conductivity is more likely to
deviate from Nernst-Einstein limit due to the cation-anion motion
correlation. As discussed earlier, this is in contrast to typical
electrolytes where Nernst-Einstein equation is often expected to be
more applicable at low ion concentrations.

\section{Conclusions}

We perform MD simulations with an applied electric field to calculate
ion conductivity directly from ion mobility. By assessing
a toy model where no electrostatic potential exists between ions,
we show that it is more accurate and efficient to calculate ion conductivity
and ion correlations from the nonequilibrium method than the typical
use of fluctuation dissipation relationships in equilibrium simulations.
After validating the accuracy, we use this method to further probe
the effects of ion-ion and ion-monomer interactions on ion transport
across a wide range of ion concentrations with our recently-developed
model (with a $1/r^{4}$ potential form to represent ion solvation).

We find that cation-anion motion is more correlated at lower concentrations
when other variables are held constant. We demonstrate that this phenomenon
is due to the slower ion cluster relaxation rate at low concentrations
rather than the static spatial state of ion aggregation or the fraction
of free ions. Two takeaways from these results are: 1) In salt-doped polymer
systems, the Nernst-Einstein equation is more applicable at higher concentrations,
which in contrast to typical electrolytes where correlation in cation-anion
motion is often expected to be reduced at low ion concentrations.
2) The number of free ions, which is often expected to be
related to ion conductivity, does not fully explain the dynamic ion
motion correlations or conductivity.

We consider systems with various ion-monomer and ion-ion solvation strengths. Optimal molar conductivity can be obtained by tuning ion-monomer
solvation strength. However, we show that the window to improve ion
conductivity by increasing dielectric constant is smaller at higher
concentrations. Stronger ion-monomer interactions causes competing
diffusion and ion correlation effects at low concentrations but significantly
reduces ion conductivity at high concentrations due to the dominant
decrease in the diffusion constant. On the other hand, increasing
ion-ion interactions impedes ion conduction regardless of ion concentration. 

Finally, at all ion-monomer and ion-ion interactions considered in
this work, ion conductivity was found well predicted by ion diffusion
at higher concentrations. On the contrary, the degree of uncorrelated
ion motion cannot be neglected to accurately calculate ion conductivity
at low concentrations. From an experimental standpoint, our results
suggest that analysis such as electrophoretic NMR (eNMR)\cite{zhang2014observation} or broadband dielectric spectroscopy (BDS)\cite{stacy2018fundamental} is needed to probe the actual correlations of ion motion rather than pulsed field gradient NMR which measures diffusion of different species. In addition, scaling conductivity or diffusion constant
based on fraction of free/dissociated ions in these systems may not be appropriate.\cite{colby1997polyelectrolyte,duluard2008lithium} By elucidating how these key factors affect
ion conductivity and correlations, we hope to provide insight on how
to optimize ion conduction in future materials.

\section{Acknowledgements}
We thank Venkat Ganesan for for helpful discussions regarding the conductivity calculation. This material is based upon work supported by the U.S. Department of Energy, Office of Science, Office of Basic Energy Sciences, under Award No. DE-SC0014209 (K.-H.S.). Additionally, this material is based upon work supported by the National Science Foundation under Grant No. 1454343 (L.M.H.). We acknowledge the support and high performance computing (HPC) resources from the Ohio Supercomputing Center.\cite{OhioSupercomputerCenter1987}

\bibliography{shen_conductivity}

\end{document}